\documentstyle[12pt,epsf]{article}

\newcommand{\be}{\begin{equation}}
\newcommand{\ee}{\end{equation}}
\newcommand{\bea}{\begin{eqnarray}}
\newcommand{\eea}{\end{eqnarray}}
\renewcommand{\appendix}{%
\renewcommand{section}{
\newpage\thispagestyle{\plain}%
\secdef\Appendix\sAppendix}%
\setcounter{section}{0}%
\renewcommand{\thesection}{\Alph{section}}}

\newcommand{\sAppendix}[1]{
{\flushright\large\bfseries\appendixname\par
\nohyphens\centering#1\par}%
\vspace{\baselineskip}}
\begin{document}
\begin{titlepage}

\begin{flushright}
\end{flushright}
\vskip 1.2cm

\begin{center}

 {\LARGE\bf Thermal Effects and Flat Direction\\ \vskip
0.2in Baryogenesis }

\vskip 1.4cm

{\large Alexey Anisimov}
\\
\vskip 0.4cm
{\it Santa Cruz Institute for Particle Physics,
     Santa Cruz, CA, USA  } \\
\vskip 0.6cm

\vskip 4pt

\vskip 1.5cm

\begin{abstract}
\hspace{0.5cm} In this paper we provide a detailed numerical study
of the influence of thermal effects on the original picture of the
Affleck-Dine (AD) baryogenesis. These effects are found to modify
the results greatly in some cases. We estimate the baryon/entropy
ratio and provide numerical results on the typical behaviour of
the charge as a function of the strength of the potential and
other parameters.
\end{abstract}
\end{center}

\vskip 1.0 cm

\end{titlepage}
\setcounter{footnote}{0} \setcounter{page}{2}
\setcounter{section}{0} \setcounter{subsection}{0}
\setcounter{subsubsection}{0}


{\LARGE{\bf 1.~Introduction}}

Affleck-Dine baryogenesis from flat directions
\cite{afd,dineetall} is a natural mechanism to explain the
baryon/entropy ratio. The current known value of $\Omega_b h^2$ is
$0.019\pm 0.002$ from nucleosynthesis and $0.031\pm 0.005$ from
BOOMERANG data \cite{boom}. This corresponds to $n_B/n_{\gamma}
\approx 5.1\times 10^{-9}$ and $n_B/n_{\gamma} \approx 8.3\times
10^{-9}$ respectively. The ingredients needed in general to
generate the baryon asymmetry are CP violating A-terms, terms
which lift flat directions, and supersymmetry breaking terms in
the early
 Universe which give rise to negative mass squared of order of $H^2$,
where $H$ is the Hubble constant. At the level of renormalizable
interactions there are several flat directions which can be found
in MSSM, i.e., directions where $F$ and $D$ terms in the potential
vanish. A simple example is the $H_uL$ flat direction. These flat
directions can  be lifted by nonrenormalizable terms which we
might imagine to be associated with the Planck scale, i.e.,
suppressed by some power of $M_*$. \footnote{By $M_*$ we denote
reduced Planck mass, $M_*=10^{18}{\rm GeV}$.} In that situation as
a result of the balance between soft induced SUSY breaking
potential and higher order nonrenormalizable terms the field can
acquire a large expectation value which evolves as some power of
$t$. Eventually, the negative mass squared term, which decreases
as $1/t^2$, becomes comparable with $m_{3/2}^2$ term and the
AD-field starts to oscillate. This is the moment when baryon
charge, which is generated by the torque due to the different
phases and time dependence of A-terms, ``freezes'' at some value.
Terms which lift the flat direction in general have the form
$\frac{\phi^{2n+4}}{M^{2n}_*}$. We will principally consider
$n=1,2,3$ and refer to these as the $n=1, 2, 3$ cases. One then
can do a simple estimate to obtain the ratio $n_B/n_{\gamma}$. For
example, in the $n=1$ case, $n_B/n_{\gamma}$
 is of order
$(m_{3/2}/M_*)^{1/2}{\rm sin(\delta)}\sim 10^{-8}{\rm sin(\delta)}$ or so.
\footnote{By $\delta$ we denote CP-violating phase.}
The resulting baryon/entropy ratio
is a bit lower because one has to take into account that, for example, in
the case of the $H_uL$ direction lepton number is produced first, which later
has to be converted to baryon number. Also, the AD condensate interacts
with thermalized inflaton decay products and some part of it can be
evaporated before the charge is produced. There is a related issue of
Q-ball formation and the corresponding evaporation rate,
 which is different from that of homogeneous condensate. There have
 been a number of papers on the subject in which
authors considered thermal effects relative to AD baryogenesis as
well as Q-ball formation. In these paper we will consider AD
baryogenesis in the absence of Q-balls. Taking into account the
formation of Q-balls will alter the whole picture. We refer the
reader to \cite{japanese} as for this subject.\\

  Recently it was noticed \cite{allaetall} that due to the evolution of the
AD condensate in the background of inflaton decay products there
is an interesting effect that takes place in addition to the
physics of the original scenario. It was observed that
superpotential interactions couple the flat directions to other
fields. These fields acquire masses induced by the flat-direction
vev but they may be sufficiently small so that fields come to
thermal equilibrium  with the inflaton decay products. In such
cases the flat direction starts to oscillate at earlier time than
usually estimated because it acquires thermal mass
$y^2T^2$\footnote{$y$ is the coupling constant and T is the
temperature.} which decreases with time as $1/t$. Since $-H^2$
falls more rapidly with time, the difference $y^2T^2-H^2$
eventually becomes positive. That normally happens much earlier
than $t\sim m^{-1}_{3/2}$.\\

It was also argued that the main source for generating baryon
asymmetry in that case are $A$-terms which are proportional to
$T$. In \cite{ad} it was pointed out that such $A$-terms are
suppressed, in general, by symmetries so that there is no
temperature enhancement of the $A$-term. However, it was noted
that there is an additional source of $A$-terms that can be
efficient in the $n=2,3$ cases. It was also shown that there is an
additional thermal contribution to the potential of the form
$T^4\log(|\phi|^2)$. In the $n=2,3$ cases, this defines the time
when the condensate starts to oscillate, rather than the $y^2 T^2$
contribution found in \cite{allaetall}.\\

 In this paper we further investigate this scenario.
We consider different types of thermal effects which are relevant
for different choices of $n$. We analyse the parameter space in
greater detail, i.e., the dependence of the resulting
baryon/entropy ratio on the parameters of the lagrangian. These
parameters are the Yukawa constant, gauge coupling constant, the
relative phase between $A$-terms, and the coefficients in front of
the $A$-terms.

The $n=1$ case is found not to generate sufficiently large
$n_B/n_{\gamma}$, but the $n=2, 3$ cases can rather easily
generate the needed number for a wide range of parameters.
Throughout we assume for simplicity that the ratio of the inflaton
mass to the reduced Planck mass is of order of $10^{-5}$, taking,
in general, the mass of inflaton $m_I$ to be $10^{13}{\rm GeV}$
and reduced Planck mass $M_*$ to be $10^{18}{\rm GeV}$. These
parameters appear in the estimate of the reheating temperature
$T_R$ as well as in the estimate of the baryon number violating
terms. Therefore, the choice of these parameters is important for
the estimate of $n_B/n_{\gamma}$. We discuss this in detail in
Section 4. \\

This paper is organized as follows: in Section 2 we discuss the
origin of two thermal effects as well as their relevance in the
$n=1, 2, 3$ cases. In Section 3 we introduce additional A-terms
which will be the sources for generating the baryon asymmetry. In
Section 4 we provide the estimates and numerical results for the
baryon/entropy ratio. In Section 5 we give some details of the
numerical simulations.\\
 \vspace{1cm}

 {\LARGE{\bf 2.~Thermal effects}}

As was argued in \cite{allaetall} and \cite{ad} there are two
types of thermal-induced contributions to the potential. The first
is due to the following mechanism. Consider some field $\chi$
which couples to the flat direction via the superpotential $W\sim
y\chi\chi\phi$ and interacts with the dilute plasma produced by
the inflaton decay. Because of $\chi$'s coupling to the flat
direction, $chi$ acquires mass $m_{\chi}=y\phi$. If this mass is
less than the temperature of the thermal plasma this field will
come to thermal equilibrium, giving an effective
temperature-dependent mass to the flat direction. Let us to do
some estimates to see when this effect might be important.
Consider the case $n=1$. Since $T=(H\Gamma_d
M_*^2)$\footnote{$\Gamma_d$ is the inflaton decay rate.} and
$\phi=(HM_*)^{1/2}$, the condition $y\phi<T$ leads to
\be
H_{th}< \frac{T_R^2}{y^4M_*}, \ee where $H_{th}$ is the value of
Hubble constant when thermalization of $\chi$ occurs. Taking $M_*$
to be $10^{18} {\rm GeV}$ and $T_R=10^{10}{\rm GeV}$ one obtains:
\be
H_{th}< (\frac{0.01}{y})^4 10^{10} {\rm Gev}. \ee On the other
hand, if $\phi$ gets a thermal mass, oscillations start when
$yT\sim H$, which gives the value of the Hubble constant $H_o$ at
this moment
\be
H_{o}\sim y^{4/3}T_R^{2/3}M_*^{1/3}=(\frac{y}{0.01})^{4/3} 10^{10}
{\rm Gev}. \ee In order for this thermal term to be relevant
oscillations have to start not later than $\chi$ becomes
thermalized, so that we have $H_o/H_{th}\leq 1$, which means
$y\leq 0.01$. Provided $y$ is sufficiently small, there is a
$y^2T^2|\phi|^2$-contribution to the potential, which affects the
time when oscillation of $\phi$ begins. If $y$ is bigger than
0.01, then oscillations start when $y^2T^2$ term is generated,
which means that $H_o$ is defined by $H_{th}$. Estimating $H_{th}$
for the $n=2$ case gives:
\be
n=2: H_{th}< \frac{T_R^6}{y^{12}M_*^5}=(\frac{0.01}{y})^{12}
10^{-6} {\rm GeV}, \ee which is too late for thermalization
process to have any effect on $V(\phi)$ unless $y\leq 0.001$. For
$n=3$ from the condition $y\phi<T$ one obtains
$(\frac{y}{0.01})<10^{-2}$. We conclude that in the $n=2, 3$ cases
this effect does not take place unless $y$ is very small.\\

  The other contribution to the potential comes from the modification of
the coupling constant when some sfermions which are coupled to the
flat direction gain masses. The effective potential for $\phi$
(see \cite{ad}) is then
\be
V_{eff}\sim \alpha^2T^4\log(|\phi|^2),
 \ee
 where $\alpha$ is the gauge coupling. This contribution
causes $\phi$ to oscillate when $H^2\sim \frac{\partial
V_{eff}}{\partial |\phi|^2}$. For $n=1$, $\phi$ starts to
oscillate at $H_o\sim \alpha T_R=(\frac{\alpha}{0.01}) 10^{8} {\rm
GeV}$. For $n=2$: $H_o\sim \alpha^{6/5}T_R
(\frac{T_R}{M_*})^{1/5}=(\frac{\alpha}{0.01})^{6/5} 10^{6} {\rm
GeV}$,  while for $n=3$: $H_o\sim \alpha^{4/3}T_R
(\frac{T_R}{M_*})^{1/3}= (\frac{\alpha}{0.01})^{4/3} 10^{5} {\rm
GeV}$. If we now look at the ratio of $H_{osc}$'s due to both
thermal contributions in the $n=1$ case, we see that
\be
\frac{H_o^{(1)}}{H_o^{(2)}}\sim \frac{y^{4/3}}{\alpha}
\frac{M_*}{T_R}=\frac{y^{4/3}}{\alpha} 10^{3}, \ee where by
$H_o^{(1)}$ and $H_o^{(2)}$ we denote the values of the Hubble
constant when the AD field starts to oscillate if there is only
$\alpha^2T^4\log(|\phi|^2)$ or $y^2T^2$ thermal contribution to
the potential, respectively.

That estimate shows that in the $n=1$ case the $y^2T^2$ term
dominates over the $\alpha^2T^4\log(|\phi|^2)$-potential unless
the ratio $\frac{y^{4/3}}{\alpha}\leq 10^{-3}$. Therefore, we will
analyze the $n=1$ case with the $y^2T^2$ term only, while for
$n=2, 3$ we will consider only the logarithmic thermal term.
\vspace{1cm}

{\LARGE{\bf 3.~$A$-terms}}

To build up a baryon or lepton asymmetry one needs to have
corresponding $U(1)$ violating $A$-terms. The misalignment between
their phases then exerts a torque, making $\phi$ to roll down
towards one of the discrete minima along the angular direction and
settle there. In the original scenario \cite{dineetall} there were
two A-terms : $Am_{3/2}\phi^{n+3}/M^n_*$, which is the usual
supersymmetry breaking $A$-term, and $aH\phi^{n+3}/M^n_*$, which
is induced because of the finite energy density in the Universe
during inflation.\footnote{$A$ and $a$ are some complex
constants.} At later times ($H < m_{3/2}$), because of its
dependence on $H$, the second term is no longer important, but at
the times $H\sim m_{3/2}$, when the flat direction starts to
oscillate, this term is of comparable size with the first one and
a sufficient amount of charge is produced.
\\

As we learned in the previous section, $\phi$-oscillations generically
start much earlier than $H\sim m_{3/2}$. One might then consider
some other sources of A-terms, which can be relevant in this case. These
additional A-terms may arise in the following way: consider the superpotential
\be
\delta W=\int d^4\theta f(I)\frac{\phi^{n+3}}{M_*^n},
 \ee
 where $f(I)$ is some holomorphic function of ithe inflaton field $I$. Taking
the first two terms in the polynomial expansion of $f(I)$ in
powers of $I/M_*$ leads to the superpotential
\be
\delta
W=\frac{1}{M_*}(aI+b\frac{I^2}{M_*})\frac{\phi^{n+3}}{M_*^n}, \ee
where, in general, $a$ and $b$ are complex constants which do not
need to have the same phase. $I$ decreases as $I_0 (t_0/t)$, so
that the second term is somewhat suppressed with respect to the
first, but at the time when oscillations of the $\phi$-field start
this is not necessarily a large suppression and one might expect
to generate a reasonable baryon number. We will investigate this
particular case in greater detail in the next section.
 \vspace{1cm}

{\LARGE{\bf 4.~$n_B/n_{\gamma}$: estimates and numerical results}}

Before we move to the discussion of numerical results let us
provide some crude estimates for the baryon number. In the $n=1$
case we assume that $y\leq0.01$ and $y^2T^2$ dominates over the
logarithmic term. Then, $H_o\sim
y^{4/3}T_R(\frac{M_*}{T_R})^{1/3}$. We take the equation
\be
\frac{dn_B}{dt}=|V_B|\sin(\delta), \ee where $\delta$ is some
effective phase which comes from the phase difference of two
$A$-terms, $V_B$ is baryon number violating part of the potential.
For the A-terms which come from superpotential $\delta
W=\frac{1}{M_*}(aI+b\frac{I^2}{M_*}) \frac{\phi^{n+3}}{M_*^n}$
$\delta$ would be the phase of $b$ after we redefine the phase of
$\phi$ to make the $aI$ term real.\footnote{The phase of $I$ is
assumed to be a constant, so that one can absorb its phase in $b$
and consider the inflaton to be real.}

Taking the initial amplitude of $I$ to be of order of $M_*$ and
replacing $\frac{dn_B}{dt}$ by the product $n_BH_o$ one obtains
\be
n_Bt^2 =
\frac{bH_o}{M_*}\frac{\phi^4}{M_*}\frac{1}{H^2_o}=by^{4/3}
T_R^{2/3}M_*^{1/3}\approx b(\frac{y}{0.01})^{4/3}10^{11} {\rm
GeV}.
 \ee
 We compute $n_Bt^2$ instead of $n_B$ because it is more
convenient, since after the oscillations begin the baryon density
decreases as  $1/t^2$. For the same reason it is more convenient
to compute $n_{\gamma}t^2= T_R^3t_d^2$ where
$t_d=(\Gamma_d)^{-1}$, and $\Gamma_d\approx m_I^3/M_*^2$ is the
inflaton decay rate. Since $T=(H\Gamma_dM_*^2)^{1/4}$, the
reheating temperature is given by $T_R=(\Gamma_dM_*)^{1/2}$. Then,
$T_R^3t_d^2=M_*^{3/2}t_d^{1/2}$. Taking $m_I\approx 10^{13}{\rm
GeV}$ one obtains $\Gamma_d\approx 10^{3} {\rm GeV}$, $T_R\approx
10^{10}{\rm GeV}$ and $n_{\gamma}t^2\approx 10^{25}{\rm GeV}$. The
value of $T_R$ is somewhat large from the perspective of the
gravitino problem. The baryon/entropy ratio is
\be
n=1 : \frac{n_B}{n_{\gamma}}\approx b(\frac{y}{0.01})^{4/3}
10^{-14} \sin({\delta_b}),
 \ee
 which turns out to be too small unless
$y\sim 1$. Actual numerical study shows that for a wide range (see
Fig. 1) of $y$ the resulting baryon/entropy ratio is somewhat
larger:
\be
\frac{n_B}{n_{\gamma}}\approx
(b\sin({\delta_b}))(10^{-14}-10^{-13}), \ee and almost independent
of $y$. There is some uncertainty which depends on the choice of
$M_*$ as well as $m_I$. However, in order for $n=1$ case to be
viable we would need to consider a much smaller ratio of $M_*/m_I$
than $10^5$.

One can also consider $\Gamma_d = \epsilon m_I^3/M_*^2$, where
$\epsilon$ is some new parameter, i.e., $\Gamma_d =
m_I^3/\Lambda^2$, where $\Lambda$ is some new scale different from
$M_*$. Then $T_R$ is lower than before by a factor of
$\epsilon^{1/2}$. For $\epsilon\sim 10^{-4}$,  $T_R\sim 10^{8}{\rm
GeV}$, which is an upper bound on  $T_R$ to avoid gravitino
problem. If one combines the result of Eq. (10) and the value for
$T_R^3t_d^2$ in terms of $\epsilon$, the inflaton mass, and $M_*$,
one gets that
$n_B/n_{\gamma}=by^{4/3}\epsilon^{5/6}(m_I/M_*)^{5/2}$. This is an
even smaller number than at $T_R=10^{10}{\rm GeV}$. So if one
tries to fit the value of $T_R$ in a consistent way to avoid
gravitino problem one obtains that the $n=1$ case seems to be even
less acceptable. In the literature sometimes the value of $T_R$ is
considered as a free parameter. In that case the value of the
ratio $\frac{n_B}{n_{\gamma}}\sim T_R^{-7/3}$ is very sensitive to
the value of the reheating temperature and our estimate for the
resulting baryon/entropy ratio changes by several orders of
magnitude. For example, if one takes $T_R=10^8{\rm GeV}$ again and
keeps $\Gamma_d\approx 10^3{\rm GeV}$ one will obtain
$\frac{n_B}{n_{\gamma}}\sim 10^{-9}$. One can also treat $t_d$ as
a free parameter as well. This, in fact, would significantly relax
constraints on $n_Bt^2$ and would soften our conclusions. There
was also the discussion in the literature \cite{japnew}, that the
gravitino problem can actually be avoided even at
 temperatures higher than $10^{10}{\rm GeV}$.
In this paper we stick to $T_R\approx 10^{10}{\rm GeV}$ according
to the estimate which we get from the expression $T=
(H\Gamma_dM_*^2)^{1/4}$ for the temperature during the inflation
at $H\approx\Gamma_d$ as well as the estimate for $\Gamma_d$
mentioned above. Therefore, for that choice of parameters, we can
conclude that $n=1$ case can hardly be acceptable. However, with
all the remarks above, one can probably consider the $n=1$ case to
be a borderline case. \\

In the case of $n=2$, a similar analysis gives $H_o\sim
\alpha^{6/5}T_R^{6/5}M_*^{-1/5}$ (now there is no $y^2T^2$-term
and oscillations start due to the logarithmic contribution). Since
$\phi$ behaves as $(HM_*^2)^{1/3}$ one obtains the following
estimate for $n_Bt^2$:
\be
n=2: n_Bt^2\approx b\alpha^{4/5}T_R^{4/5}M_*^{1/5}\approx
b(\frac{\alpha}{0.1})^{4/5} 10^{11} {\rm GeV},
 \ee
 and the
corresponding baryon/entropy ratio is
\be
n=2 : \frac{n_B}{n_{\gamma}}\approx b(\frac{\alpha}{0.1})^{4/5}
10^{-13} \sin(\delta_b). \ee This is very different from the
numerical result, which is
\be
\frac{n_B}{n_{\gamma}}\approx (b\sin(\delta_b))
(10^{-10}-10^{-9}),
 \ee
 for wide range of ${\alpha}$ (see Figs.2, 3). We
want to point out here that as in the case $n=1$ the dependence of
the result on the coupling constant is rather weak, as opposed to
what one gets estimating $n_B/n_{\gamma}$ analytically.

Before we explain why the numerical result  is so different from
the naive estimate, we want to repeat the analysis above for the
$n = 3$ case. I that case $\phi = (HM_*^3)^{1/4}$; the operator
which creates the charge is $\frac{bH^2}{M_*}
\frac{\phi^6}{M_*^3}$. Estimating $n_B t^2$ at $H=H_o$ one gets
$n_B t^2 \approx bH_o^{1/2} M_*^{1/2}\sin(\delta_b)$. Since
$H_o\approx \alpha^{6/5} T_R^{6/5}/M_*^{1/5}$,
\be
n_B t^2 \approx b\alpha^{2/3}\sin(\delta_b) T_R^{2/3}M_*^{1/3}.
 \ee
 Taking $T_R=10^{10} {\rm GeV}$ and $M_*=10^{18}{\rm GeV}$ one
gets
\be
n_B t^2 \approx b\alpha^{2/3}\sin(\delta_b) 10^{12} {\rm GeV},
 \ee
 and, the corresponding baryon/entropy ratio
\be
\frac{n_B}{n_{\gamma}}=\frac{n_Bt^2}{T_R^3t_d^2}\approx
b\alpha^{4/5}\sin(\delta_b) 10^{-13},
 \ee
 which is too small for
all reasonable values of $b$, $y$ and $\sin(\delta_b)$. One can
notice that this rough estimate predicts the decrease of
baryon/entropy ratio as $y$ gets smaller as well as in the $n=2$
case. Actual numerical study, as in case $n=2$, gives a quite
different value of that ratio. The baryon/entropy ratio for the
wide range of $\alpha$ is
\be
\frac{n_B}{n_{\gamma}}\approx  b\sin(\delta_b)(10^{-8}-10^{-9}),
 \ee
 which could be consistent with experimentally known
value if $b\sim (0.1-0.01)$ and $\delta_b\sim 1$.\\

The naive expectation fails because the approximation of $dn_B/dt$
by $n_B H_o$ is too crude with respect to actual integration. One
can understand why the amplitude of $n_Bt^2$ grows as $y$ gets
smaller, when looking at the behavior of $n_Bt^2$ at some values
of $y$ as a function of time. From numerical results it is clear
that the later the oscillations of $\phi$ start the more
oscillations $n_B t^2$ undergoes before it ``freezes''. With each
oscillation, the operator which is responsible for the charge
production contributes more and more to the amplitude of $n_B
t^2$. That might explain the unexpected behavior of charge in the
$n=2, 3$ cases. One can see that the behavior of the charge for
$n=1$ with the $y^2T^2$-term is quite different from its behavior
for $n=2, 3$ with the logarithmic term. Namely, for $n=2, 3$,
before the value of a charge ``freezes'', it experiences many more
oscillations than in the $n=1$ case, gaining larger amplitude with
each of them. One can see this from numerical
 results in Fig.1-6, which show the evolution of the charge
with time and with the strength of the potential (Yukawa coupling
in the case $n=1$, and gauge coupling in the $n=2, 3$ cases). The
first three plots demonstrate the expected behavior of the charge,
i.e., it oscillates at the time $t<t_o$ and freezes afterwards.
The difference between $n=1$ and $n=2, 3$ is that for $n=1$ the
charge typically oscillates one or two times, while in the $n=2,
3$ cases it oscillates much more. Correspondingly, for $n=1$, the
numerical result is close to the naive estimate, but for $n=2, 3$
numerical value is much larger.\\

 The Figs.1-3 further
distinguish these cases. One can see that changing the Yukawa
coupling in the phenomenologically interesting region  does not
produce major changes in the amplitude of $n_Bt^2$ in the $n=1$
case. However, if one changes the value of the gauge coupling in
the $n=2, 3$ cases one can see that $n_Bt^2$ oscillates with
$\log(\alpha)$ and its amplitude grows with the decreasing of
$\alpha$, which is against the naive analytical estimate. We
believe that this happens due to the logarithmic nature of the
thermal potential in these cases.   \\

Dependence of numerical result of the baryon/entropy ratio on $b$
and $\delta_b$ doesn't bring any surprises with respect to the
crude estimate above. In Fig.7 one can see that dependence on
$\delta_b$ is $\sin(\delta_b)$, and from  Fig.8 the dependence on
$b$ is a linear function. Thus for any $b$ and $\delta_b$ one
knows how to reproduce the baryon/entropy ratio from the results
in the equations (12, 15, 19).
 \vspace{1cm}

{\LARGE{\bf 5.~Details of the numerical simulations}}

The equation for the evolution of the $\phi$ is
\be
\ddot{\phi} + 3H\dot{\phi} + \frac{\partial V}{\partial \phi^*} =
0. \ee In general $V$ has the form $V=V_0(|\phi|^2)+V_B(\phi)
+V_B(\phi^*)$, where $V_0$ determines the evolution of $|\phi|$
and $V_B$ is the part of the potential which is responsible for
the creating of the baryon charge. It also has a minor effect on
evolution of $|\phi|$, which can be neglected in the cases we
consider. The potentials we wish to study have the form: \bea
&V_0=(-H^2 + m^2_{3/2})|\phi|^2 + V_{th}(T) + |\frac{\partial
W}{\partial \phi}|^2,& \\ \nonumber
&V_B=aHW+b\frac{H^2}{M_*}W+Am_{3/2}W&
\eea
 As we discussed before,
the terms which are proportional to $m_{3/2}$ are not important.
Dropping them, for $n=1$ we have:
\be
V=(-H^2 + y^2T^2)|\phi|^2  + \lambda^2 \frac{|\phi|^6}{M_*^2} +
(aH\frac{\phi^4}{M_*} +b\frac{H^2}{M_*}\frac{\phi^4}{M_*} + {\rm
h.c.}),
 \ee
 for $n=2$:
\be
V=-H^2|\phi|^2  + \alpha^2T^4\log(|\phi|^2) + \lambda^2 \frac{|\phi|^8}{M_*^4}
+ (aH\frac{\phi^5}{M_*^2} +b\frac{H^2}{M_*}\frac{\phi^5}{M_*^2}
 + {\rm h.c.}),
\ee
and for $n=3$:
\be
V=-H^2|\phi|^2  + \alpha^2T^4\log(|\phi|^2) + \lambda^2 \frac{|\phi|^{10}}{M_*^6}
+ (aH\frac{\phi^6}{M_*^3} +b\frac{H^2}{M_*}\frac{\phi^6}{M_*^3}
 + {\rm h.c.}).
 \ee
 We took into account that different thermal effects are
relevant for $n=1$ and $n=2,3$. Introducing radial and angular
parts of $\phi$ by $\phi=Re^{i\Omega}$ one gets corresponding
equations for $R$ and $\Omega$:
\newline
$n=1$: \bea &\ddot{R} +3H\dot{R} + (-H^2 -
\dot{\Omega}^2+y^2T^2+\frac{\lambda^2R^4}{M^2_*})R = 0,& \\
\nonumber &\ddot{\Omega} +(3H +\frac{2\dot{R}}{R})\dot{\Omega} =
|a|\frac{R^2H}{M_*}\sin(4\Omega) +
|b|\frac{R^2H^2}{M_*^2}\sin(4\Omega-\delta_b);&
 \eea
 $n=2$: \bea
&\ddot{R} +3H\dot{R} + (-H^2 - \dot{\Omega}^2+\alpha^2T^4/R^2+
\frac{\lambda^2R^6}{M^4_*})R = 0,& \\ \nonumber &\ddot{\Omega}
+(3H +\frac{2\dot{R}}{R})\dot{\Omega} =
|a|\frac{R^3H}{M_*^2}\sin(5\Omega) +
|b|\frac{R^3H^2}{M_*^3}\sin(5\Omega-\delta_b);&
 \eea
 $n=3$: \bea
&\ddot{R} +3H\dot{R} + (-H^2 - \dot{\Omega}^2+\alpha^2T^4/R^2+
\frac{\lambda^2R^8}{M^6_*})R = 0,& \\ \nonumber &\ddot{\Omega}
+(3H+\frac{2\dot{R}}{R})\dot{\Omega} =
|a|\frac{R^4H}{M_*^3}\sin(6\Omega) +
|b|\frac{R^4H^2}{M_*^4}\sin(6\Omega-\delta_b).&
 \eea
 We are
dropping the contribution from $V_B$ at $t<t_o$ to the equations
for $R$ for the following reasons. Let us to compare the behavior
of $R$ induced by the $\frac{\lambda^2R^4}{M^2_*}$ and the
$|a|\frac{R^2H}{M_*}\cos(4\Omega)$ terms separately ($n=1$ case).
We know that the first term causes $R$ to behave as
$(HM_*)^{1/2}$. The second term causes $R$ to behave in the same
way, provided $\Omega$ is small, so that $\cos(4\Omega)\approx 1$.
The smallness of $\Omega$ is guaranteed by its equation of
evolution and, in fact, is well seen in numerical results.
Therefore, qualitatively it does not bring any changes to the
evolution of $R$.\footnote{In fact, one can also neglect
$\dot\Omega^2$ term in the equation for $R$ since this term is
small compared to the others for any $H>H_o$. It becomes, however,
important after the oscillations start. At this moment baryon
charge is already produced and fixed. Later evolution of the
baryon number is not interesting because of that and one does not
need to run simulations further.} The other term, which is
proportional to $|b|$, is suppressed by $(m_I/M_*)$ and falls down
more rapidly with time with respect to the $A$-term. Hence,
neglecting these terms in the equation for $R$ makes sense, since
that does not change the evolution of $R$ and simplifies the
numerical and analytical analysis of the equations.\\

For numerical simulations it is more convenient to introduce a
dimensionless field $r=Rt$ and scale everything by $m_I$, which we
take to be $10^{13} {\rm GeV}$. Then, the equations above with all
the simplifications which were discussed before take the form:\\
$n=1$:
 \bea
  &\ddot{r} + (-H^2 +y^2T^2+\epsilon^2
\frac{\lambda^2r^4}{t^4})r = 0,& \\ \nonumber &\ddot{\Omega}
+\frac{2\dot{r}}{r}\dot{\Omega} = |a|\epsilon
\frac{r^2H}{t^2}\sin(4\Omega) + |b|\epsilon^2
\frac{r^2H^2}{t^2}\sin(4\Omega-\delta_b);&
 \eea
 $n=2$: \bea
&\ddot{r} + (-H^2 +\alpha^2t^2T^4/r^2+
\epsilon^4\frac{\lambda^2r^6}{t^6})r = 0,& \\ \nonumber
&\ddot{\Omega} +\frac{2\dot{r}}{r}\dot{\Omega} =
|a|\epsilon^2\frac{r^3H}{t^3}\sin(5\Omega) +
|b|\epsilon^3\frac{r^3H^2}{t^3}\sin(5\Omega-\delta_b);&
 \eea
$n=3$:
 \bea
  &\ddot{r} + (-H^2 +\alpha^2t^2T^4/r^2+
\epsilon^6\frac{\lambda^2r^8}{t^8})r = 0,& \\ \nonumber
&\ddot{\Omega} +\frac{2\dot{r}}{r}\dot{\Omega} =
|a|\epsilon^3\frac{r^4H}{t^4}\sin(6\Omega) +
|b|\epsilon^4\frac{r^4H^2}{t^4}\sin(6\Omega-\delta_b),&
 \eea
  where
$\epsilon=M_I/M_*\approx 10^{-5}$. Solving these equations
numericaly and noting that $n_Bt^2=r^2\dot{\Omega}$, we can plot
the behavior of $n_Bt^2$ as a function of time as well as find its
asymptotic value as a function of different parameters. \\

At this point it is rather easy to analyze the behavior of the
charge. In fact, the equation for $r$ is no longer dependent on
$\Omega$, so that the second equation describes the evolution of
$\Omega$ in the ``background'' of the radial component, which, in
turn, is given by the first equation in each case. The evolution
of $r$ is easy to understand just looking at its equation. First,
it falls down as some power of $t$, which depends on $n$, and then
at $H\sim H_o$ starts to oscillate. Numerical comparison of the
full system of equations without all these simplifications which
we made above shows that at $t<t_o$ this approximation is
reliable.
 \vspace{1cm}

{\LARGE{\bf 6.~Conclusions}}

Clearly thermal effects are important in the evolution of AD
condensate and modify the original mechanism of AD baryogenesis.
We find that in the $n=1$ case it is difficult to reproduce the
known value of the baryon/entropy ratio even if taking into
account some uncertainty due to the choice of the value of the
reheating temperature, so that this case should probably be
considered borderline at best. However, the $n=2, 3$ cases give
plausible results for a wide range of the parameters of the
potential. We have shown that the naive estimate of
$n_B/n_{\gamma}$ fails by several orders of magnitude to reproduce
the observed numerical value. This effect is most clearly seen in
the $n=2, 3$ cases, when the radial component of the AD field
evolves in the logarithmic potential. This happens due to the
behavior of the charge before it freezes, which might seem a bit
surprising. Instead of decreasing with $H_o$ it actually slightly
grows. We have also shown numerically that dependence on a few
other parameters of the potential such as the phase difference
between the $A$ terms and the coefficient of the $A$ term $b$ is
of the expected form.\\

{\bf Acknowledgements.}

We would like to thank M. Dine and M. Graesser for useful
conversations and for the discussion of the results of this paper.



\pagebreak
 {\large
 {\LARGE{\bf 7.~Figures and Captions}}\\

Fig.1: This graph illustrates the behavior of the asymptotic value
of $n_Bt^2$ as a function of $\ln(y)$ for $n=1$. The choice of
parameters here is $a=b=1$; $\delta_b=\pi/3$.

Fig.2: This graph illustrates the behavior of the asymptotic value
of $n_Bt^2$ as a function of $\ln(\alpha)$ for $n=2$. The choice
of parameters here is $a=b=1$; $\delta_b=\pi/3$.

Fig.3: This graph illustrates the behavior of the asymptotic value
of $n_Bt^2$ as a function of $\ln(\alpha)$ for $n=3$. The choice
of parameters here is $a=b=1$; $\delta_b=\pi/3$.

Fig.4 This graph illustrates generic behavior of $n_Bt^2$ as a
function of time in the $n=1$ case; cases a, b, c, d correspond to
$y_a=0.01$, $y_b=0.005$, $y_c=0.0025$, $y_d=0.001$. The values of
other parameters are taken to be $\delta_b=\pi/3, a=b=1=\lambda$.

  Fig.5: This graph illustrates generic behavior
of $n_Bt^2$ as a function of time in the $n=2$ case; cases a, b, c
correspond to $\alpha_a=0.1$, $\alpha_b=0.07$, $\alpha_c=0.03$.
The values of other parameters are taken to be $\delta_b=\pi/3,
a=b=1=\lambda$.

Fig.6: This graph illustrates the behavior of $n_Bt^2$ for $n=3$,
case $a$ corresponds to $\alpha=0.1$, case $b$ to $\alpha=0.7$,
and case $c$ to $\alpha=0.5$. The values of other parameters are
taken to be $\delta_b=\pi/3, a=b=1=\lambda$.

Fig.7: This graph illustrates the behavior of the asymptotic value
of $n_Bt^2$ as a function of $sin(\delta_b)$ for $n=2$. The choice
of parameters here is $a=b=1$; $\alpha$=0.05.

Fig.8: This graph illustrates the behavior of charge as a function
of $b$} {\large for $n=2$. The choice of parameters here is $a=1$,
$\delta_b=\pi/3$, $\alpha$=0.03

\pagebreak

\begin{figure}[ht]
\begin{center}
\epsfxsize= 2.19 in
\leavevmode
\epsfbox[200 304 401 666]{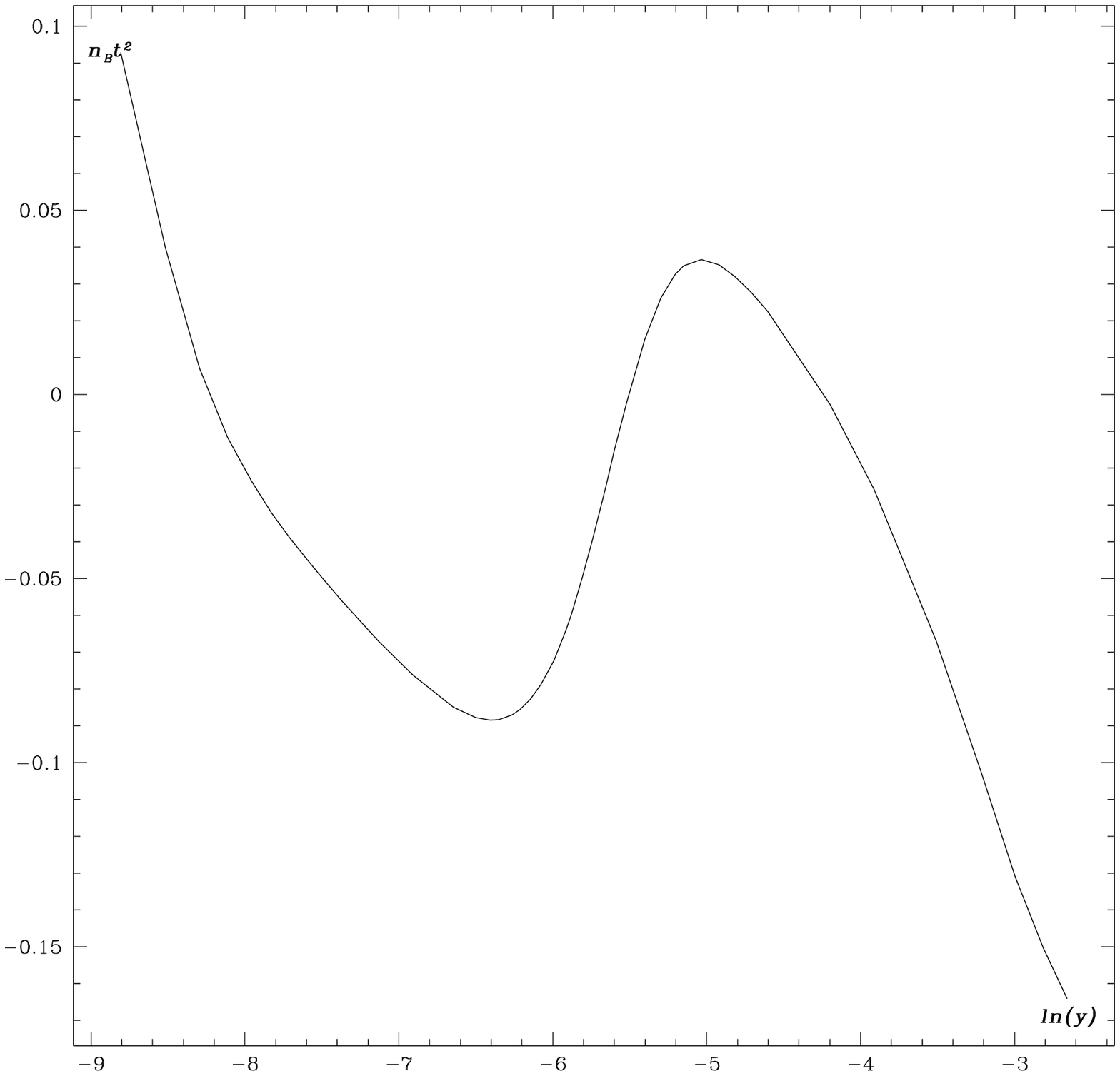}
\end{center}
\vskip 1.2in
 \caption[f1]{}
\end{figure}

\pagebreak

\begin{figure}[ht]
\begin{center}
\epsfxsize= 2.19 in
\leavevmode
\epsfbox[200 304 401 666]{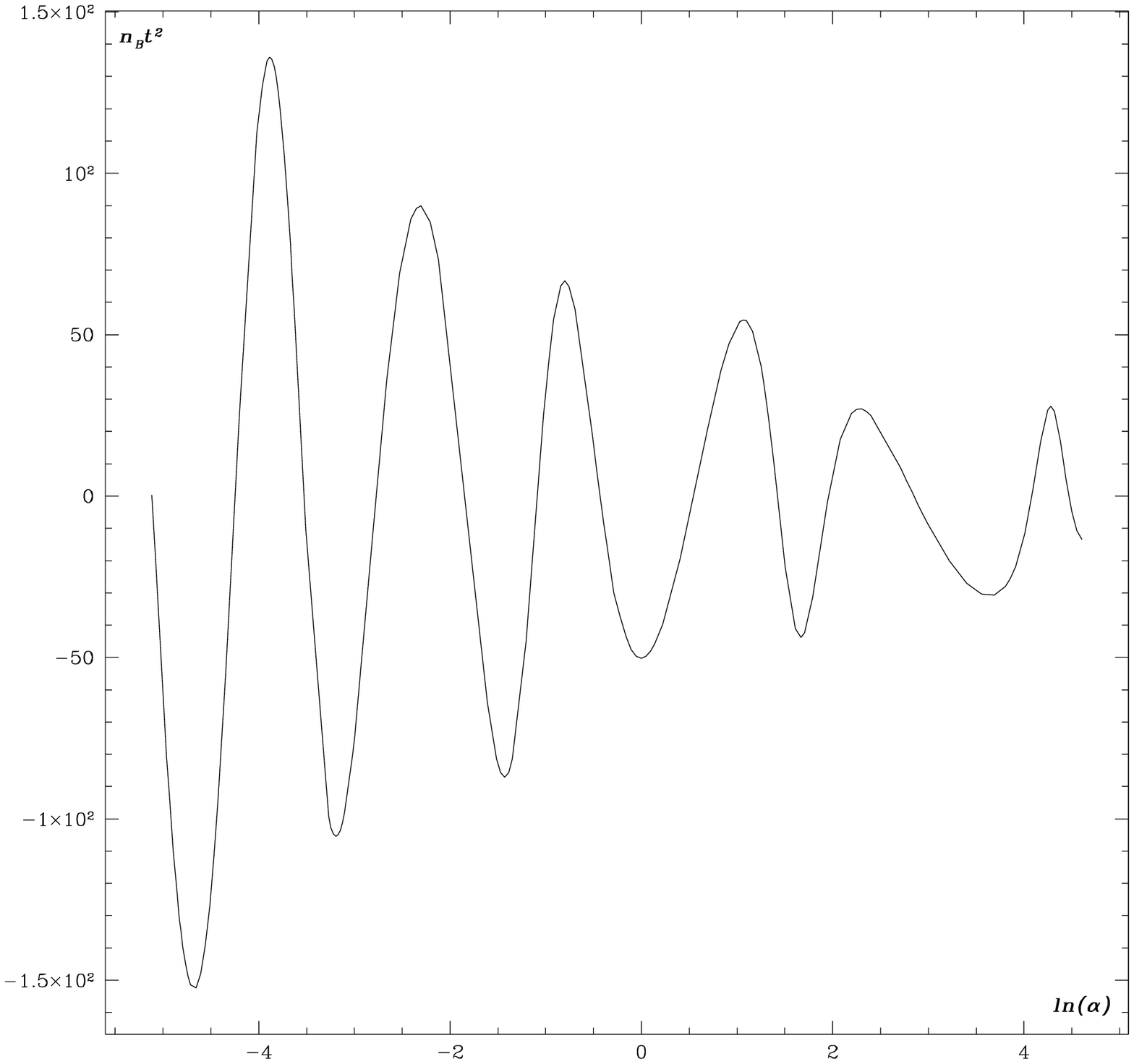}
\end{center}
\vskip 1.2in
 \caption[f2]{}
\end{figure}

\pagebreak

\begin{figure}[ht]
\begin{center}
\epsfxsize= 2.19 in
\leavevmode
\epsfbox[200 304 401 666]{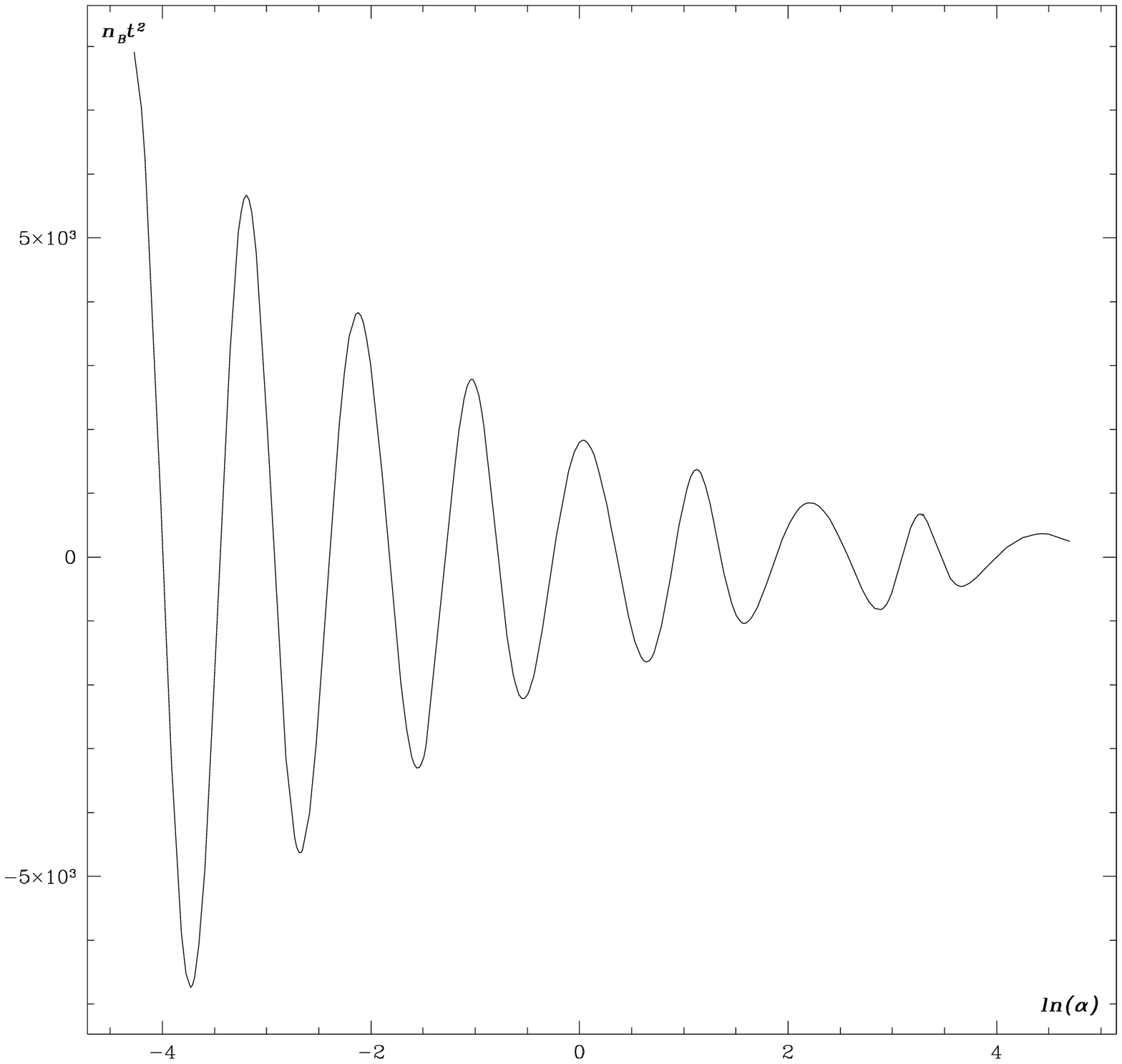}
\end{center}
\vskip 1.2in
 \caption[f3]{}
\end{figure}

\pagebreak

\begin{figure}[ht]
\begin{center}
\epsfxsize= 2.19 in \leavevmode \epsfbox[200 304 401
666]{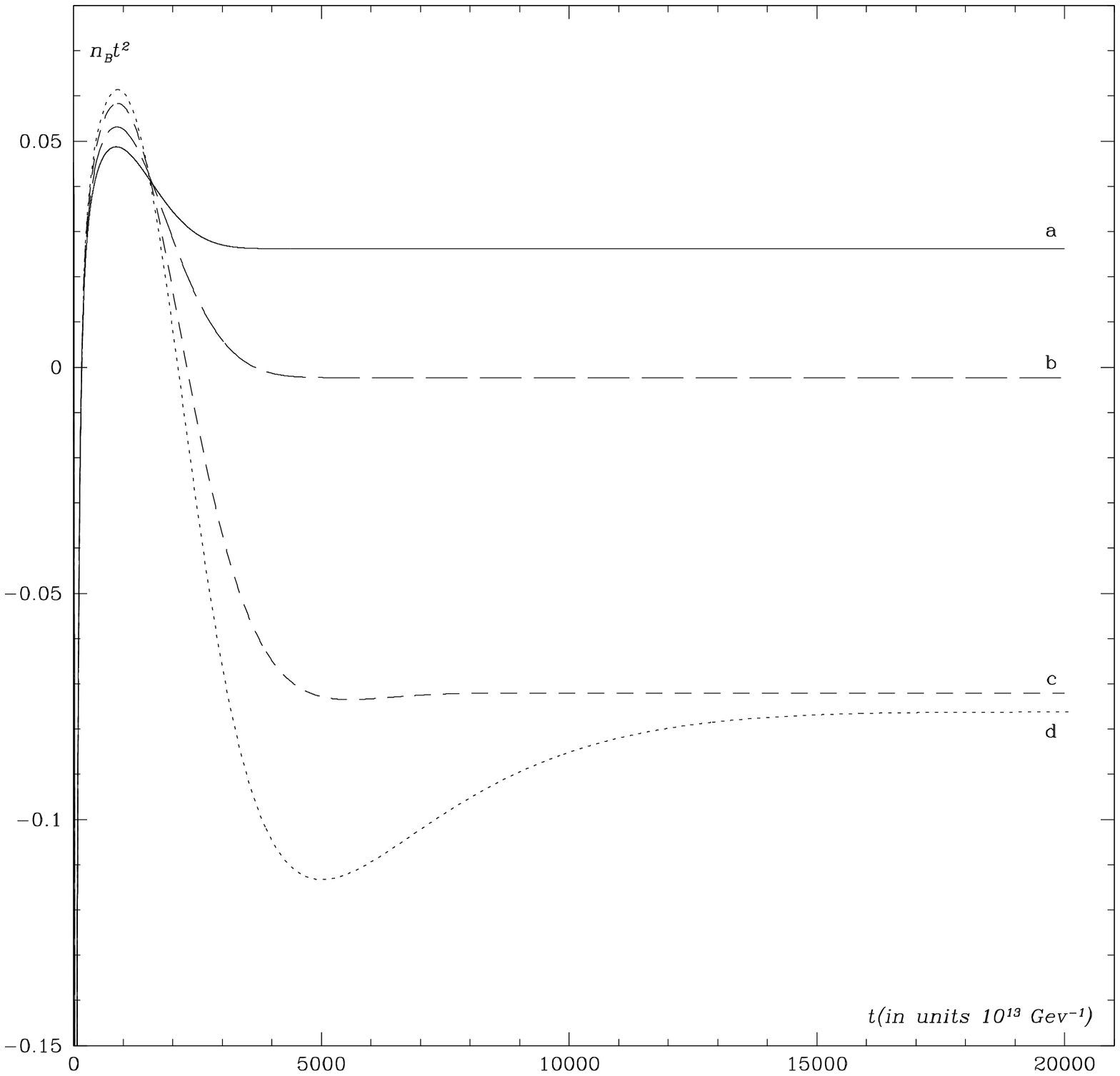}
\end{center}
\vskip 1.2in
 \caption[f4]{}
\end{figure}

\pagebreak

\begin{figure}[ht]
\begin{center}
\epsfxsize= 2.19 in \leavevmode \epsfbox[200 304 401
666]{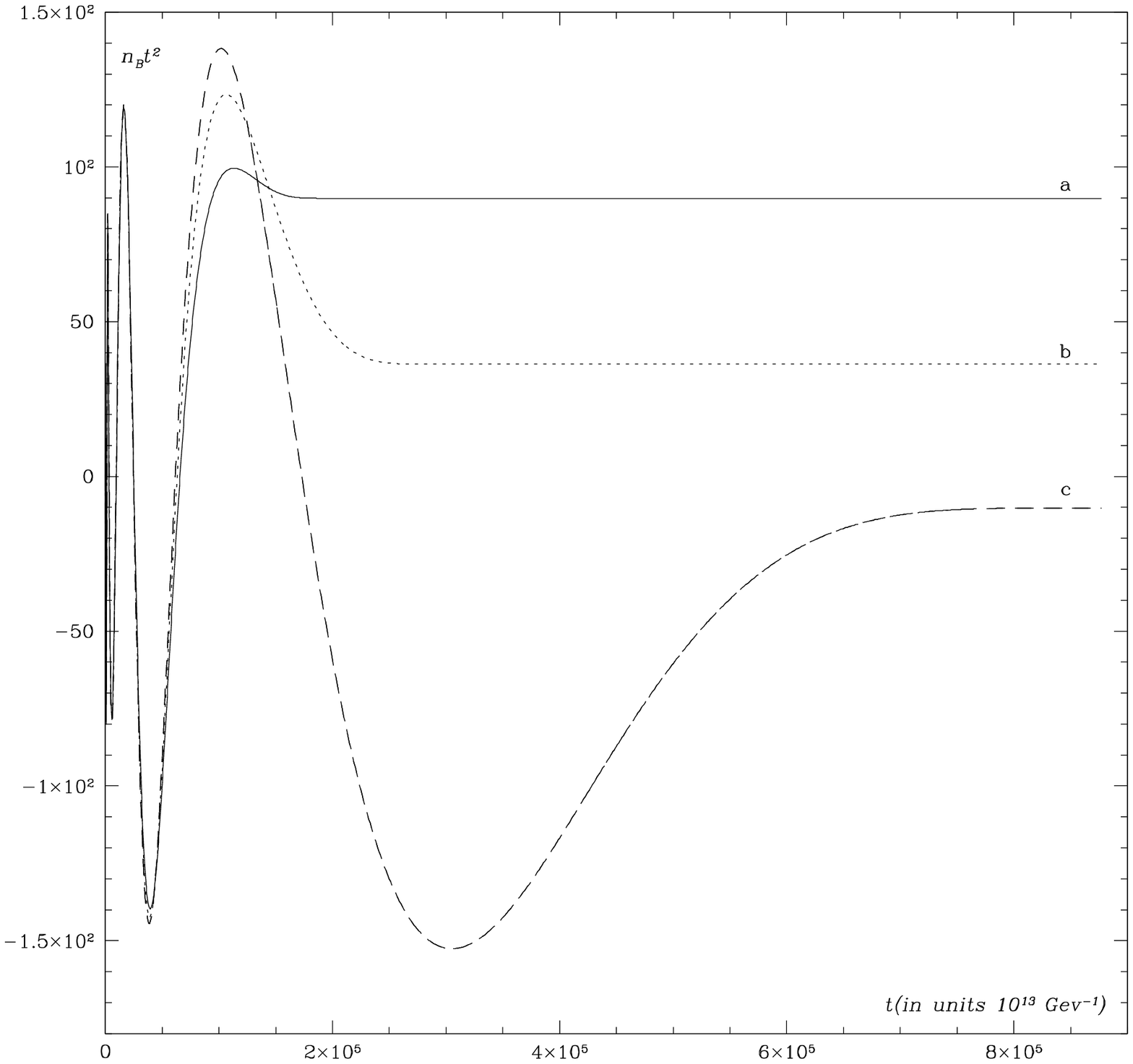}
\end{center}
\vskip 1.2in
 \caption[f5]{}
\end{figure}

\pagebreak

\begin{figure}[ht]
\begin{center}
\epsfxsize= 2.19 in \leavevmode \epsfbox[200 304 401
666]{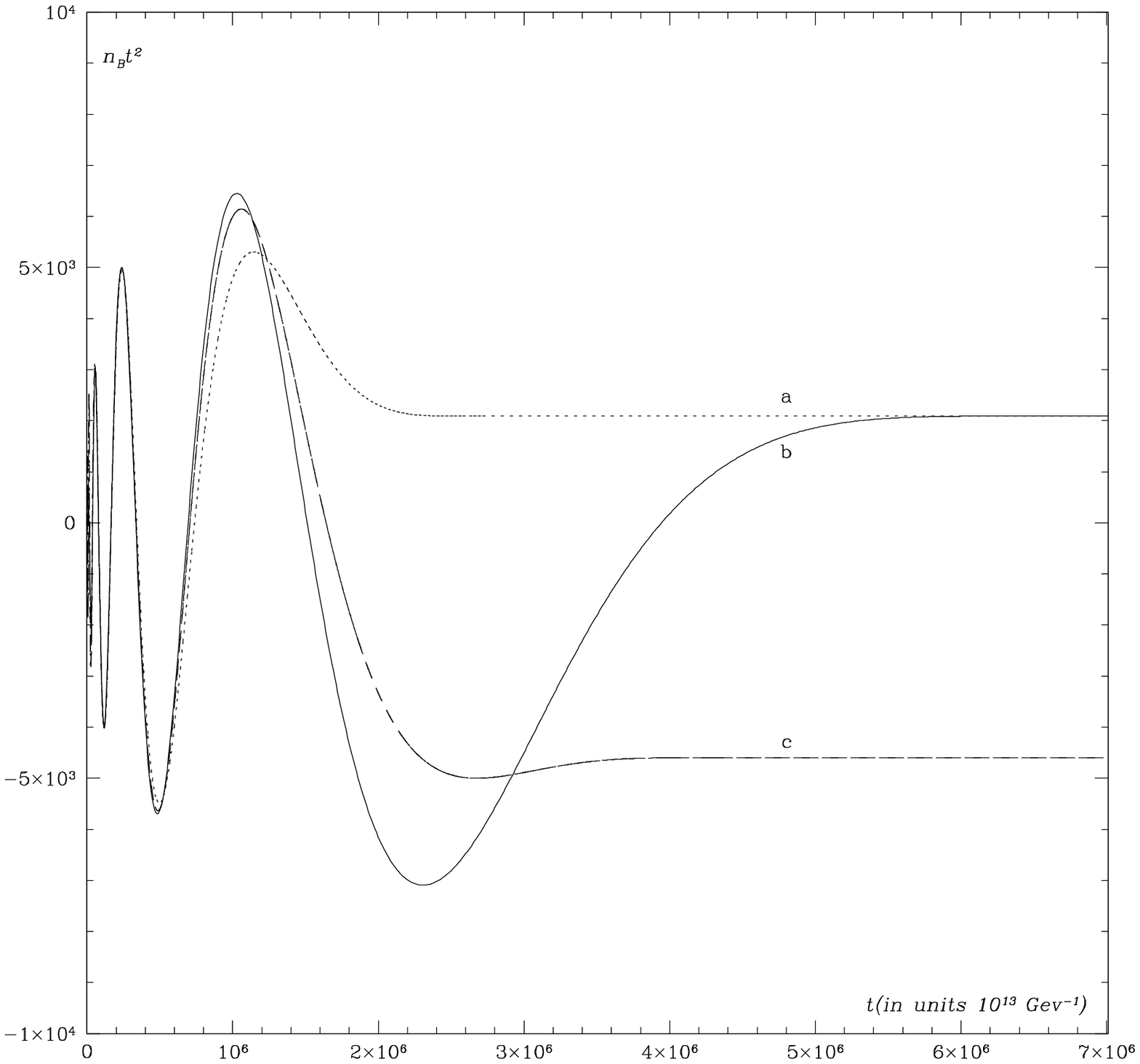}
\end{center}
\vskip 1.2in
 \caption[f6]{}
\end{figure}

\begin{figure}[ht]
\begin{center}
\epsfxsize= 2.19 in
\leavevmode
\epsfbox[200 304 401 666]{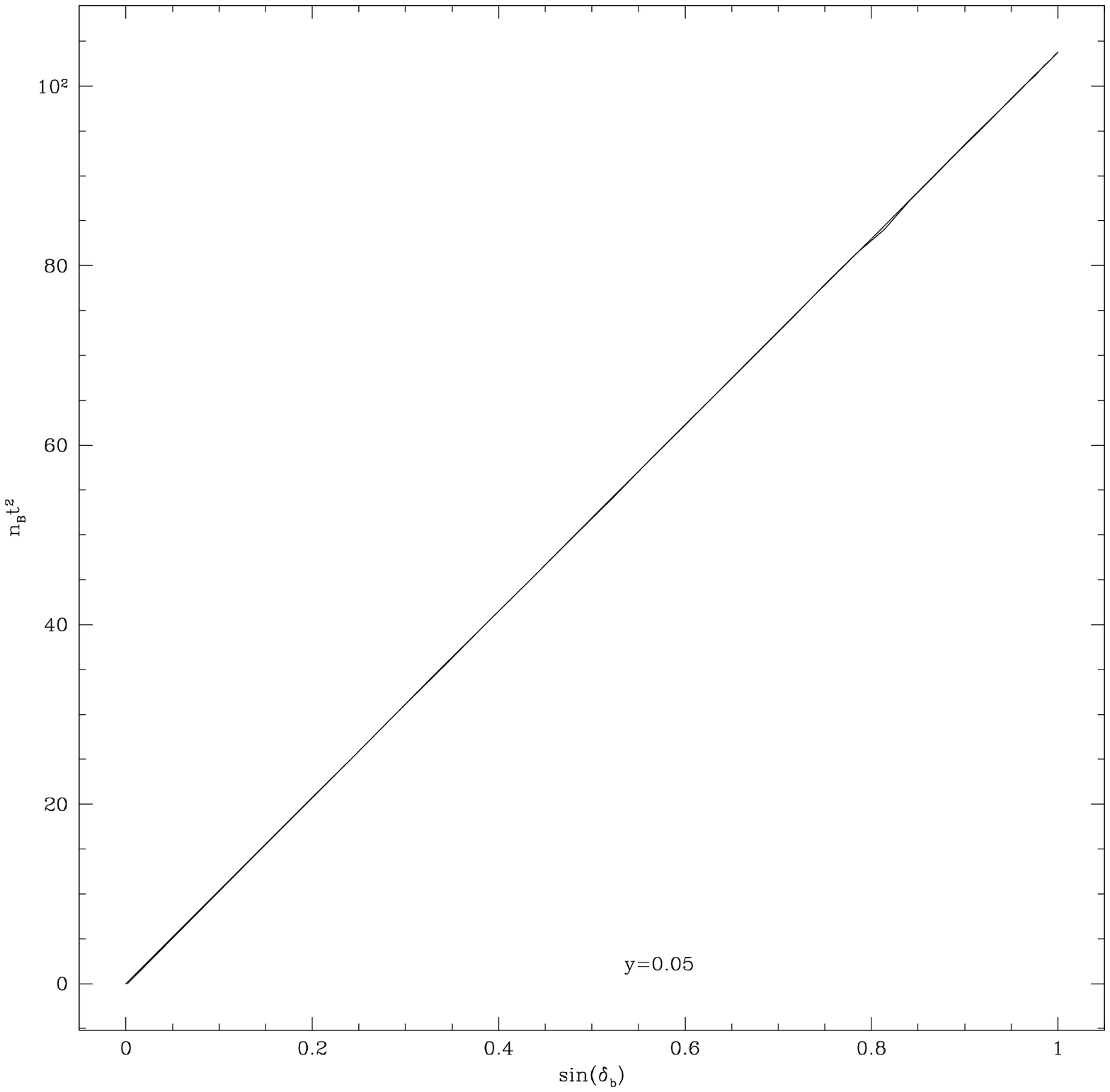}
\end{center}
\vskip 1.2in
 \caption[f7]{}
\end{figure}

\pagebreak

\begin{figure}[ht]
\begin{center}
\epsfxsize= 2.19 in
\leavevmode
\epsfbox[200 304 401 666]{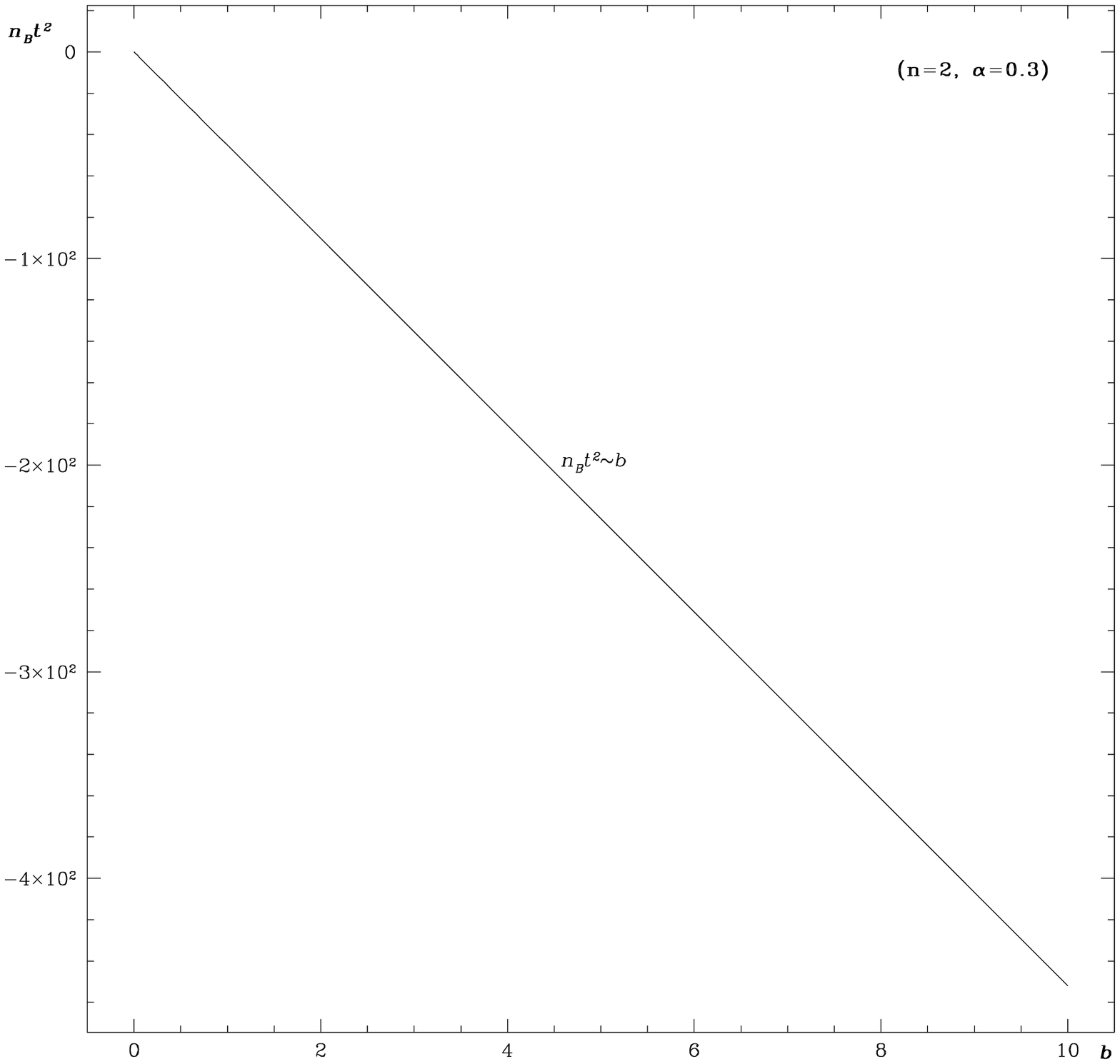}
\end{center}
\vskip 1.2in \caption[f8]{}
\end{figure}
\end{document}